\documentstyle[twoside,epsf]{article}

\catcode`\@=11
\long\def\@makefntext#1{
\protect\noindent \hbox to 3.2pt {\hskip-.9pt  
$^{{\eightrm\@thefnmark}}$\hfil}#1\hfill}		

\def\thefootnote{\fnsymbol{footnote}}
\def\@makefnmark{\hbox to 0pt{$^{\@thefnmark}$\hss}}	
        
\def\ps@myheadings{\let\@mkboth\@gobbletwo
\def\@oddhead{\hbox{}
\rightmark\hfil\eightrm\thepage}   
\def\@oddfoot{}\def\@evenhead{\eightrm\thepage\hfil
\leftmark\hbox{}}\def\@evenfoot{}
\def\sectionmark##1{}\def\subsectionmark##1{}}

\oddsidemargin=\evensidemargin
\renewcommand{\thefootnote}{\fnsymbol{footnote}}

\newcounter{sectionc}\newcounter{subsectionc}\newcounter{subsubsectionc}
\renewcommand{\section}[1] {\vspace{12pt}\addtocounter{sectionc}{1} 
\setcounter{subsectionc}{0}\setcounter{subsubsectionc}{0}\noindent 
        {\tenbf\thesectionc. #1}\par\vspace{5pt}}
\renewcommand{\subsection}[1] {\vspace{12pt}\addtocounter{subsectionc}{1} 
        \setcounter{subsubsectionc}{0}\noindent 
        {\bf\thesectionc.\thesubsectionc. {\kern1pt \bfit #1}}\par\vspace{5pt}}
\renewcommand{\subsubsection}[1] {\vspace{12pt}\addtocounter{subsubsectionc}{1}
        \noindent{\tenrm\thesectionc.\thesubsectionc.\thesubsubsectionc.
        {\kern1pt \tenit #1}}\par\vspace{5pt}}
\newcommand{\nonumsection}[1] {\vspace{12pt}\noindent{\tenbf #1}
        \par\vspace{5pt}}

\newcounter{appendixc}
\newcounter{subappendixc}[appendixc]
\newcounter{subsubappendixc}[subappendixc]
\renewcommand{\thesubappendixc}{\Alph{appendixc}.\arabic{subappendixc}}
\renewcommand{\thesubsubappendixc}
	{\Alph{appendixc}.\arabic{subappendixc}.\arabic{subsubappendixc}}

\renewcommand{\appendix}[1] {\vspace{12pt}
        \refstepcounter{appendixc}
        \setcounter{figure}{0}
        \setcounter{table}{0}
        \setcounter{lemma}{0}
        \setcounter{theorem}{0}
        \setcounter{corollary}{0}
        \setcounter{definition}{0}
        \setcounter{equation}{0}
        \renewcommand{\thefigure}{\Alph{appendixc}.\arabic{figure}}
        \renewcommand{\thetable}{\Alph{appendixc}.\arabic{table}}
        \renewcommand{\theappendixc}{\Alph{appendixc}}
        \renewcommand{\thelemma}{\Alph{appendixc}.\arabic{lemma}}
        \renewcommand{\thetheorem}{\Alph{appendixc}.\arabic{theorem}}
        \renewcommand{\thedefinition}{\Alph{appendixc}.\arabic{definition}}
        \renewcommand{\thecorollary}{\Alph{appendixc}.\arabic{corollary}}
        \renewcommand{\theequation}{\Alph{appendixc}.\arabic{equation}}
        \noindent{\tenbf Appendix \theappendixc #1}\par\vspace{5pt}}
\newcommand{\subappendix}[1] {\vspace{12pt}
        \refstepcounter{subappendixc}
        \noindent{\bf Appendix \thesubappendixc. {\kern1pt \bfit #1}}
	\par\vspace{5pt}}
\newcommand{\subsubappendix}[1] {\vspace{12pt}
        \refstepcounter{subsubappendixc}
        \noindent{\rm Appendix \thesubsubappendixc. {\kern1pt \tenit #1}}
	\par\vspace{5pt}}

\newcommand{\textlineskip}{\baselineskip=13pt}
\newcommand{\smalllineskip}{\baselineskip=10pt}

\def\eightcirc{
\begin{picture}(0,0)
\put(4.4,1.8){\circle{6.5}}
\end{picture}}
\def\eightcopyright{\eightcirc\kern2.7pt\hbox{\eightrm c}} 

\newcommand{\copyrightheading}[1]
        {\vspace*{-2.5cm}\smalllineskip{\flushleft
        {\footnotesize International Journal of Modern Physics C, #1}\\
        {\footnotesize $\eightcopyright$\,\,\, World Scientific Publishing
         Company}\\
         }}

\def\abstracts#1#2#3{{
        \centering{\begin{minipage}{4.5in}\baselineskip=10pt\footnotesize
        \parindent=0pt #1\par
        \parindent=15pt #2\par
        \parindent=15pt #3\par
        \end{minipage}}\par}} 

\renewenvironment{thebibliography}[1]
        {\frenchspacing
	 \ninerm\baselineskip=11pt
         \begin{list}{\arabic{enumi}.}
        {\usecounter{enumi}\setlength{\parsep}{0pt}     
         \setlength{\leftmargin 17pt}{\rightmargin 0pt}   
         \setlength{\itemsep}{0pt} \settowidth
	{\labelwidth}{#1.}\sloppy}}{\end{list}}

\newcounter{itemlistc}
\newcounter{romanlistc}
\newcounter{alphlistc}
\newcounter{arabiclistc}

\newcommand{\fcaption}[1]{
        \refstepcounter{figure}
	\setbox\@tempboxa = \hbox{\footnotesize Fig.~\thefigure. #1}
	\ifdim \wd\@tempboxa > 5in
           {\begin{center}
	\parbox{5in}{\footnotesize\smalllineskip Fig.~\thefigure. #1}
            \end{center}}
        \else
             {\begin{center}
	     {\footnotesize Fig.~\thefigure. #1}
              \end{center}}
        \fi}

\newcommand{\tcaption}[1]{
        \refstepcounter{table}
	\setbox\@tempboxa = \hbox{\footnotesize Table~\thetable. #1}
        \ifdim \wd\@tempboxa > 5in
           {\begin{center}
         \parbox{5in}{\footnotesize\smalllineskip Table~\thetable. #1}
            \end{center}}
        \else
             {\begin{center}
	     {\footnotesize Table~\thetable. #1}
              \end{center}}
        \fi}

\def\@citex[#1]#2{\if@filesw\immediate\write\@auxout
	{\string\citation{#2}}\fi
\def\@citea{}\@cite{\@for\@citeb:=#2\do
	{\@citea\def\@citea{,}\@ifundefined
	{b@\@citeb}{{\bf ?}\@warning
	{Citation `\@citeb' on page \thepage \space undefined}}
	{\csname b@\@citeb\endcsname}}}{#1}}

\newif\if@cghi
\def\cite{\@cghitrue\@ifnextchar [{\@tempswatrue
	\@citex}{\@tempswafalse\@citex[]}}
\def\citelow{\@cghifalse\@ifnextchar [{\@tempswatrue
	\@citex}{\@tempswafalse\@citex[]}}
\def\@cite#1#2{{$\null^{#1}$\if@tempswa\typeout
	{IJCGA warning: optional citation argument 
	ignored: `#2'} \fi}}

\def\pmb#1{\setbox0=\hbox{#1}
        \kern-.025em\copy0\kern-\wd0
        \kern.05em\copy0\kern-\wd0
        \kern-.025em\raise.0433em\box0}

\def\fnt#1#2{\footnotetext{\kern-.3em
        {$^{\mbox{\scriptsize #1}}$}{#2}}}

\def\fpage#1{\begingroup
\voffset=.3in
\thispagestyle{empty}\begin{table}[b]\centerline{\footnotesize #1}
        \end{table}\endgroup}

\def\runninghead#1#2{\pagestyle{myheadings}
\markboth{{\protect\footnotesize\it{\quad #1}}\hfill}
{\hfill{\protect\footnotesize\it{#2\quad}}}}
\font\tenbf=cmbx10
\font\tenit=cmti10 
\font\tenit=cmti10
\font\bfit=cmbxti10 at 10pt

\font\ninerm=cmr9

\font\eightrm=cmr8

\def\lsym{\raise-3pt\hbox{\vbox{\tabskip0pt\offinterlineskip
	\halign{\tabskip0pt plus 1em
	##\tabskip0pt\cr
	$\,\,<\,\,$\cr
	$\,\,\sim\,\,$\cr}}}}
\def\rsym{\raise-3pt\hbox{\vbox{\tabskip0pt\offinterlineskip
     \halign{\tabskip0pt plus 1em
      ##\tabskip0pt\cr
      $\,\,>\,\,$\cr
      $\,\,\sim\,\,$\cr}}}}
\def\qed{\hbox{${\vcenter{\vbox{			
	\hrule height 0.4pt\hbox{\vrule width 0.4pt height 6pt
	\kern5pt\vrule width 0.4pt}\hrule height 0.4pt}}}$}}
\def\theequation{\thesection.\arabic{equation}}		
\renewcommand{\thefootnote}{\fnsymbol{footnote}}	

\begin{document}

\runninghead{H. Kawamura \& K. Hukushima}
{Dynamical aspects of a three-dimensional Heisenberg spin glass}

\normalsize\textlineskip
\thispagestyle{empty}
\setcounter{page}{1}

\copyrightheading{Vol. 0, No. 0 (2000) 000--000}

\vspace*{0.88truein}

\fpage{1}
\centerline{\bf  Dynamical aspects of equilibrium and off-equilibrium 
simulations }
\vspace*{0.035truein}
\centerline{\bf of a three-dimensional Heisenberg spin glass}
\vspace*{0.37truein}
\centerline{\footnotesize H. Kawamura} 
\vspace*{0.015truein}
\centerline{\footnotesize\it 
Kyoto Institute of Technology, Sakyo-ku, Kyoto 606-8585, Japan}
\vspace*{0.15truein}
\centerline{\footnotesize and}
\vspace*{0.15truein}
\centerline{\footnotesize K. Hukushima}
\vspace*{0.015truein}
\centerline{\footnotesize\it 
ISSP, University of Tokyo, Minato-ku, Tokyo 106-8666, Japan}

\vspace*{0.225truein}

\vspace*{0.21truein}
\abstracts{
Spin-glass and chiral-glass orderings of a three-dimensional isotropic
Heisenberg
spin glass are studied both by equilibrium and off-equilibrium Monte
Carlo simulations with emphasis on their dynamical aspects. 
The  model is found to exhibit a finite-temperature
chiral-glass transition without the conventional spin-glass
order. Although chirality is an Ising-like quantity from symmetry,
universality class of the chiral-glass transition appears to be
different from that of the standard Ising spin glass. In the
off-equilibrium simulation, while the spin autocorrelation exhibits only 
an interrupted aging, the chirality autocorrelation persists to exhibit
a pronounced aging effect reminiscent of the one observered in the
mean-field model. 
}{}{}



\setcounter{section}{1}
\setcounter{equation}{0}
\vspace*{1pt}\textlineskip	
\section{Introduction}		
\vspace*{-0.5pt}
\noindent
Ordering of complex systems has 
attracted  interest of 
researchers working in the field of
numerical simulations.
Well-known examples of such complex systems 
may be a variety of glassy systems including window glasses,
orientational glasses of molecular crystals, 
vortex glasses in superconductors and
spin-glass magnets.
Often, in the
dynamics of such complex systems,  characteristic slow relaxation 
is known to occur.
It has been a great challenge for  researchers   in the field
to clarify the nature and the origin of these slow dynamics, as
well as to
get fully equilibrium properties by overcoming
the slow relaxation.
In particular, {\it spin glasses\/} 
are the most extensively studied typical
model system, 
for which numerous analytical, numerical and experimental
works have been made.\cite{Reviews}

Studies on spin glasses now have more than twenty years of history.
Main focus of earlier studies was put on obtaining the
equilibrium properties of spin glasses.
While extensive experimental studies have clarified that 
the spin-glass magnets exhibit an equilibrium  phase transition 
at a finite temperature,\cite{Reviews} 
the true nature of the experimentally observed spin-glass transition and
that of the low-temperature spin-glass phase still remain open problems. 
It has been known that the magnetic interactions in many of 
real spin-glass materials are Heisenberg-like, in the sense that the
magnetic anisotropy is much weaker than the exchange energy.
In apparent contrast to experiments, numerical simulations have
indicated that isotropic Heisenberg spin glass exhibits only a
zero-temperature
transition.\cite{BanavarCieplak,McMillan,OYS,MII,Kawamura92,Kawamura95,Kawamura96} 
Apparently, there is a puzzle here: How can one reconcile the
absence of the spin-glass order in an isotropic 
Heisenberg spin glass with the experimental observation? 

In order to solve this apparent puzzle, a  chirality mechanism of
experimentally observed spin-glass transitions
was recently proposed by one of the authors,\cite{Kawamura92,Kawamura96} 
on the assumption 
that an isotropic 3D Heisenberg
spin glass 
exhibited a finite-temperature 
{\it chiral-glass\/} transition without the conventional spin-glass 
order, in which only spin-reflection 
symmetry was broken with
preserving spin-rotation symmetry. ``Chirality'' is an Ising-like
multispin variable representing the
sense or the handedness of the noncoplanar spin structures. 
It was argued that, in real spin-glass magnets, 
the spin and the chirality were ``mixed'' due to the weak
magnetic anisotropy and the chiral-glass transition
was then ``revealed'' via anomaly in
experimentally accessible quantities. 
Meanwhile, 
theoretical question whether there really occurs such 
finite-temperature chiral-glass transition in an isotropic 3D
Heisenberg spin glass,
a crucial assumption of the chirality mechanism, 
remains somewhat inconclusive.\cite{Kawamura96,HK}

More recently, 
there arose a growing interest both theoretically and
experimentally in the {\it off-equilibrium\/} dynamical properties of 
spin glasses. In particular,
aging phenomena observed
in many spin glasses\cite{LSNB}
have attracted  attention of researchers.\cite{VHOBC,BCKM}
Unlike systems in thermal equilibrium,  relaxation of physical
quantities depends not only on the observation time $t$ but also on 
the waiting time $t_w$, {\it i.e.\/}, how long
one waits at a given state before the measurements. 
Recent studies  have revealed that
the off-equilibrium dynamics  in the spin-glass  state generally 
has two characteristic time regimes.\cite{VHOBC,BCKM,CK}  
One is a short-time regime, $t_0\ll t\ll t_w$ ($t_0$ is a microscopic
time scale),  called ``quasi-equilibrium regime'', and the
other is a long-time regime, $t\gg t_w$, called ``aging regime'' or
``out-of-equilibrium regime''. In the quasi-equilibrium regime,
the relaxation is stationary and the fluctuation-dissipation
theorem (FDT) holds. 
On the other hand, in the aging regime, the relaxation becomes
non-stationary, FDT broken. 
Although theoretical studies of off-equilibrium dynamics on Ising-like
models\cite{Rieger,CK,Baldassarri,TYH} 
succeeded in reproducing some of the features of experimental results,
it is clearly desirable to study the dynamical properties of a {\it
Heisenberg} spin-glass model to make a direct link between theory
and experiments. 

In the present article, we report on our recent results of 
equilibrium as well as off-equilibrium Monte Carlo (MC) simulations on
an isotropic 3D Heisenberg spin glass, with emphasis on their dynamical
aspects. 
Ordering properties of both
the spin and the chirality are studied, aimed at 
testing the
validity of the proposed 
chirality scenario of spin-glass transitions.
We note that MC simulation is particularly suited to this
purpose,
since, at the moment, chirality itself is not directly 
measurable experimentally. By contrast,  
it is 
straightforward to measure the chirality by numerical simulations.

\textheight=7.7truein		
\setcounter{footnote}{0}
\renewcommand{\thefootnote}{\alph{footnote}}

\setcounter{section}{2}
\setcounter{equation}{0}
\section{Model}
\noindent
The model we simulate  is the classical Heisenberg
model on 
a simple cubic lattice 
with the nearest-neighbor  random Gaussian couplings, $J_{ij}$,  
defined by the Hamiltonian
\begin{equation}
{\cal H}=-\sum_{<ij>} 
J_{ij}{\mit\bf S}_i\cdot{\mit\bf S}_j
,
\label{eqn:model}
\end{equation}
where ${\bf S}_i = (S_i^x,S_i^y,S_i^z)$ is a three-component unit
vector, and the sum runs over all nearest-neighbor pairs 
with $N=L\times L\times L$ spins.   
$J_{ij}$ is the isotropic exchange coupling 
with zero mean and variance $J$. 
 
Frustration in vector spin systems often causes {\it noncollinear\/} or
{\it noncoplanar\/} spin structures. Such noncollinear or noncoplanar
orderings give rise to a nontrivial chirality according as the spin
structure is either right- or left-handed.\cite{Villain}
We define the local chirality at the $i$-th site and in the $\mu $-th 
direction, 
$\chi _{i\mu }$, for three Heisenberg spins by,
\begin{equation}
\chi _{i\mu }={\bf S}_{i+\hat {\bf e}_\mu }\cdot ({\bf S}_i\times 
{\bf S}_{i-\hat {\bf e}_\mu }),
\end{equation}
where $\hat {\bf e}_\mu \ (\mu =x,y,z)$ 
denotes a unit lattice vector along the
$\mu\/$-axis.
The chirality is a pseudoscalar in the sense that it is invariant under
global spin rotation but changes sign under global spin reflection.
Chiral order is related to a possible breaking of the spin-reflection
symmetry with preserving the spin-rotation symmetry. 

\setcounter{section}{3}
\setcounter{equation}{0}
\bigskip
\section{Equilibrium simulations}
\noindent
First, we report on our {\it equilibrium\/} MC simulations
of a fully isotropic 3D Heisenberg spin glass defined by
Eq.~(\ref{eqn:model}).\cite{HK}
Monte Carlo simulation is performed based on an exchange method
developed by Hukushima and Nemoto,\cite{HN} where whole configurations
at two neighboring temperatures of the same sample are occasionally
exchanged  keeping the system at equilibrium. 
The method has turned out to be efficient in
thermalizing the system showing slow relaxation over free energy 
barriers.\cite{HN} 
By this method, we 
succeed in equilibrating the system down to the temperature
considerably lower than those attained in the previous simulations.
We run  in parallel two independent replicas 
with the same bond realization
and compute an overlap between the chiral variables in the two
replicas,
\begin{equation}
q_{\chi } =  {1\over 3N}\sum _{i,\mu} \chi _{i\mu} ^{(1)}
\chi _{i\mu} ^{(2)}.
\end{equation}

The lattice sizes studied are
$L=6,8,10,12,16$ with periodic boundary conditions.
In the case of $L=16$, for example, we prepare 60
temperature points distributed in the range [$0.10J, 0.25J$] 
for a given sample, 
and perform $1.2\times 10^6$
exchanges  per temperature of the whole lattices
combined with the same number of standard 
single-spin-flip heat-bath sweeps.
For $L=16$, we equilibrate the system down to
the temperature $T/J=0.10$, 
which is lower than the minimum
temperature attained previously.
Sample average is taken over 1500 ($L=6$), 1200 ($L=8$), 640 ($L=10$),
296 ($L=12$) and 136 ($L=16$) independent bond realizations.
Equilibration is checked by monitoring the 
stability of the results
against at least three-times longer runs for a subset of samples.

\begin{figure}
\epsfxsize=0.49\columnwidth\epsfbox{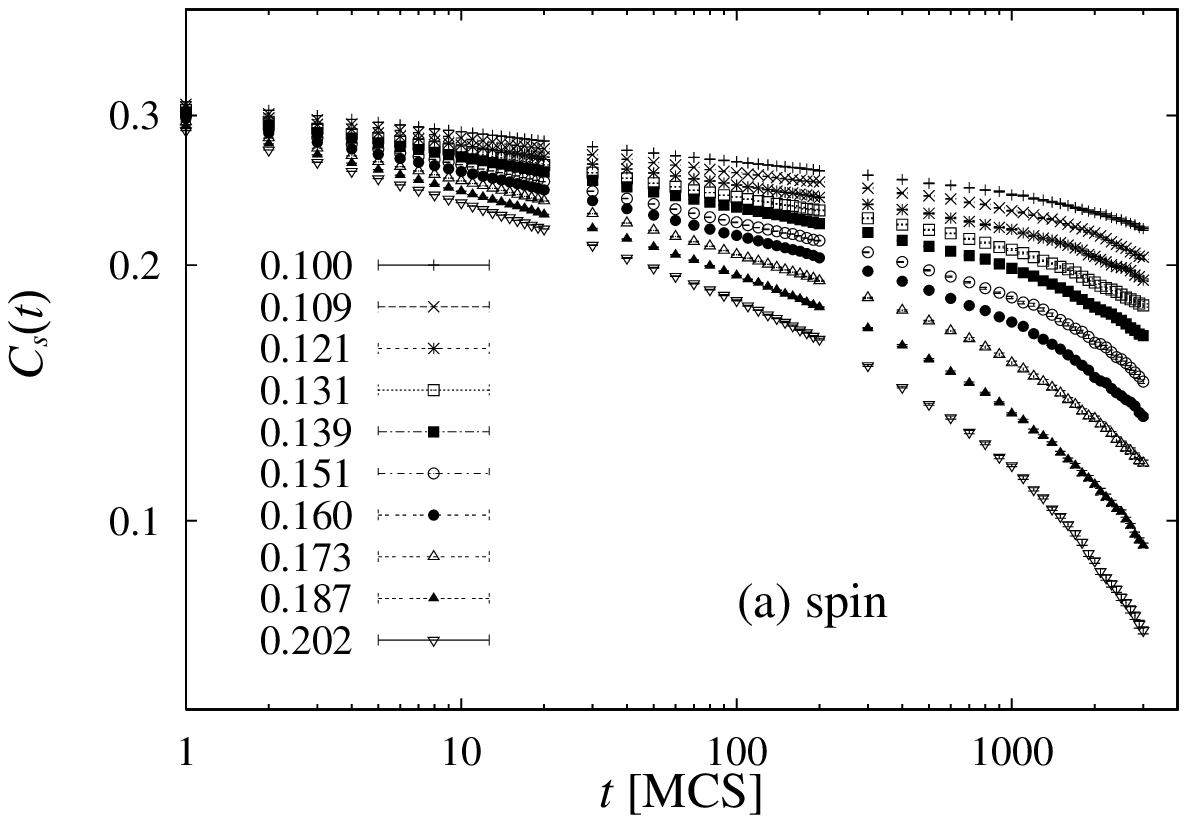}
\epsfxsize=0.49\columnwidth\epsfbox{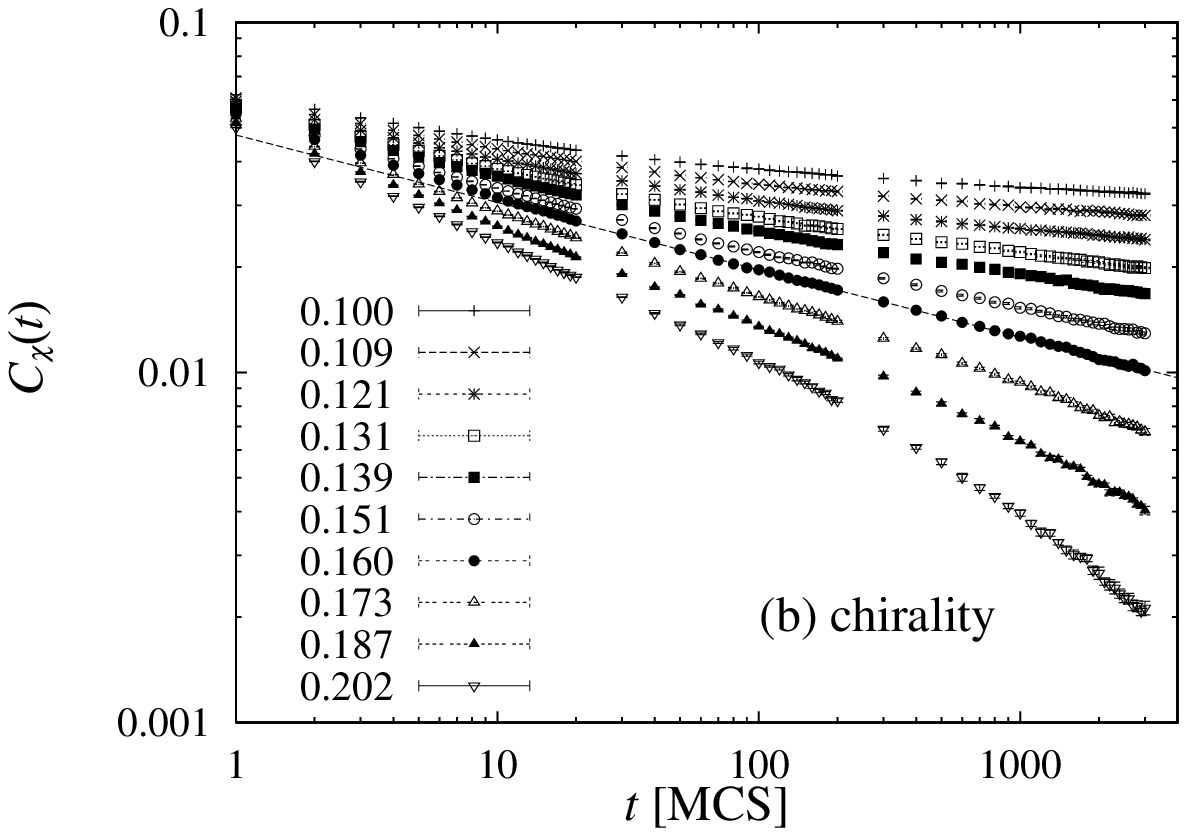}
\vspace*{10pt}
\fcaption{
Log-log plots of the 
time dependence of the equilibrium
spin (a) and chirality (b) autocorrelation 
functions 
at several temperatures.
The lattice size
is $L=16$ averaged over 64 samples.
}
\label{fig:coft-eq}
\end{figure}

Information  about the equilibrium dynamics 
can be obtained via
the spin and chirality autocorrelation 
functions defined by
\begin{eqnarray}
C_s(t) & = & \frac{1}{N}
\sum _i[\langle {\mit\bf  S}_i(t_0)\cdot {\mit\bf S}_i(t+t_0)\rangle],\\
C_\chi (t) & = & \frac{1}{3N}\sum _{i,\mu }[\langle \chi _{i\mu }(t_0) 
\chi _{i\mu }(t+t_0)\rangle],
\end{eqnarray}
where $\langle\cdots\rangle$ and $[\cdots ]$ represent
the thermal average and 
the average over bond disorder, respectively,
while $t_0$ denotes some initial time at which the system is
already fully equilibrated.
Although we have
used the exchange MC method for thermalizing the system, the time
evolution during the measurements of 
autocorrelations has been made via the
standard single-spin-flip heat-bath updating.
MC time dependence of the
calculated $C_s(t)$ and $C_\chi (t)$ 
are shown in Figs.~\ref{fig:coft-eq} on log-log plots
for several temperatures.
As can be seen from Fig.~\ref{fig:coft-eq}(a), $C_s(t)$ 
shows a downward curvature at all
temperature studied, suggesting an exponential-like decay
characteristic of the disordered phase, consistently with the
absence of the standard spin-glass order. In sharp contrast to this,
$C_\chi (t)$ shows either a downward curvature 
characteristic of the disordered phase, or an upward curvature
characteristic of the long-range ordered phase, 
depending on whether the temperature is higher or lower than
$T/J\simeq 0.16$, 
while just at $T/J\simeq 0.16$ the linear behavior corresponding to
the power-law decay 
is observed: See Fig.~\ref{fig:coft-eq}(b).
Hence, our dynamical data indicates that the chiral-glass
order without the standard spin-glass order 
takes place at $T_{{\rm CG}}/J=0.160\pm 0.005$,  
below which a finite chiral
Edwards-Anderson (EA) order parameter $q_{{\rm CG}}^{{\rm EA}}>0$
develops. 
From the slope of the data at $T=T_{{\rm CG}}$,
the exponent $\lambda $ characterizing the power-law decay of 
$C_\chi (t)\approx t^{-\lambda }$ is estimated to be
$\lambda =0.193\pm 0.005$.
We note that the occurrence of a chiral-glass transition was also 
supported by the
behaviors of the static Binder ratios (not shown here).\cite{HK}

\begin{figure}[h]
\centerline{
\epsfxsize=0.5\columnwidth\epsfbox{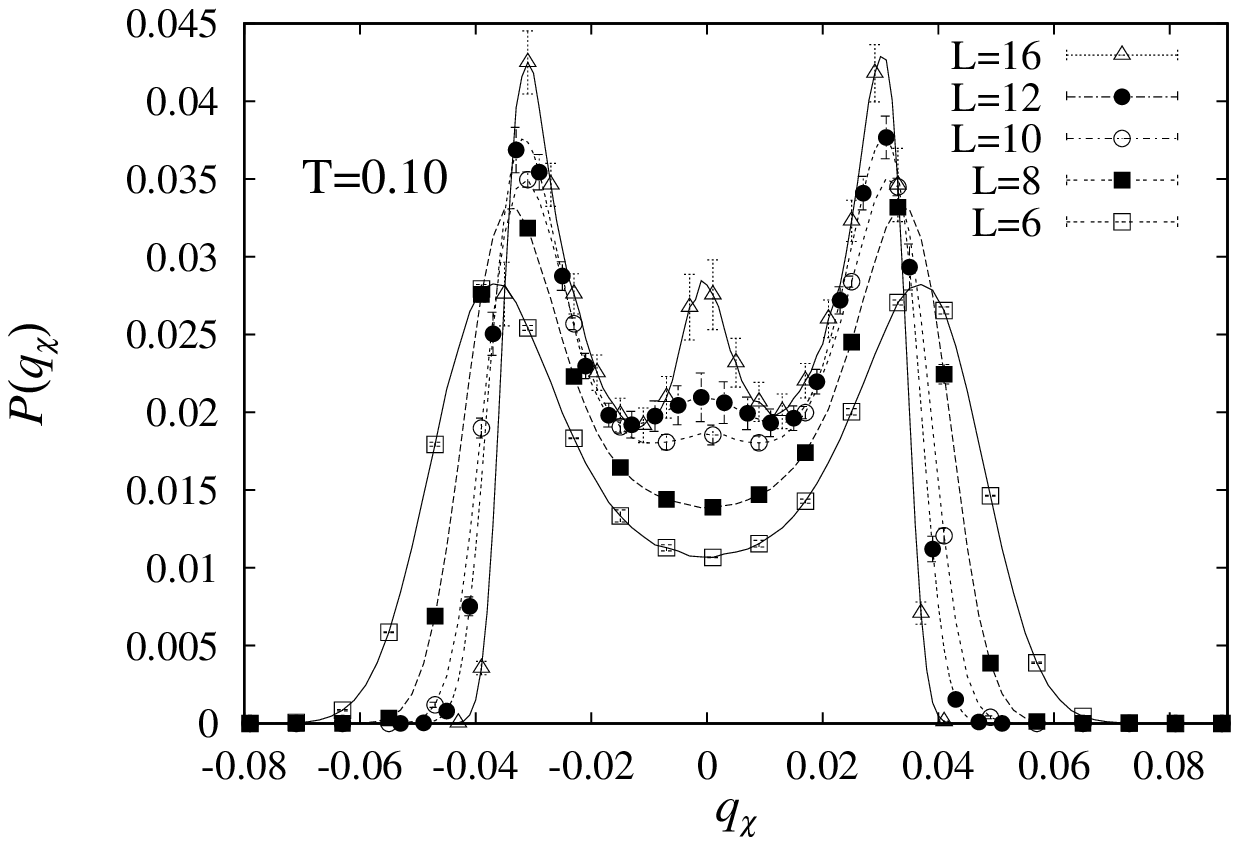}
}
\vspace*{10pt}
\fcaption{
Chiral-overlap distribution function 
below $T_{{\rm CG}}$. The temperature is 
$T/J=0.1$.
}
\label{fig:pofq}
\end{figure}

\begin{figure}[h]
\centerline{
\epsfxsize=0.48\columnwidth\epsfbox{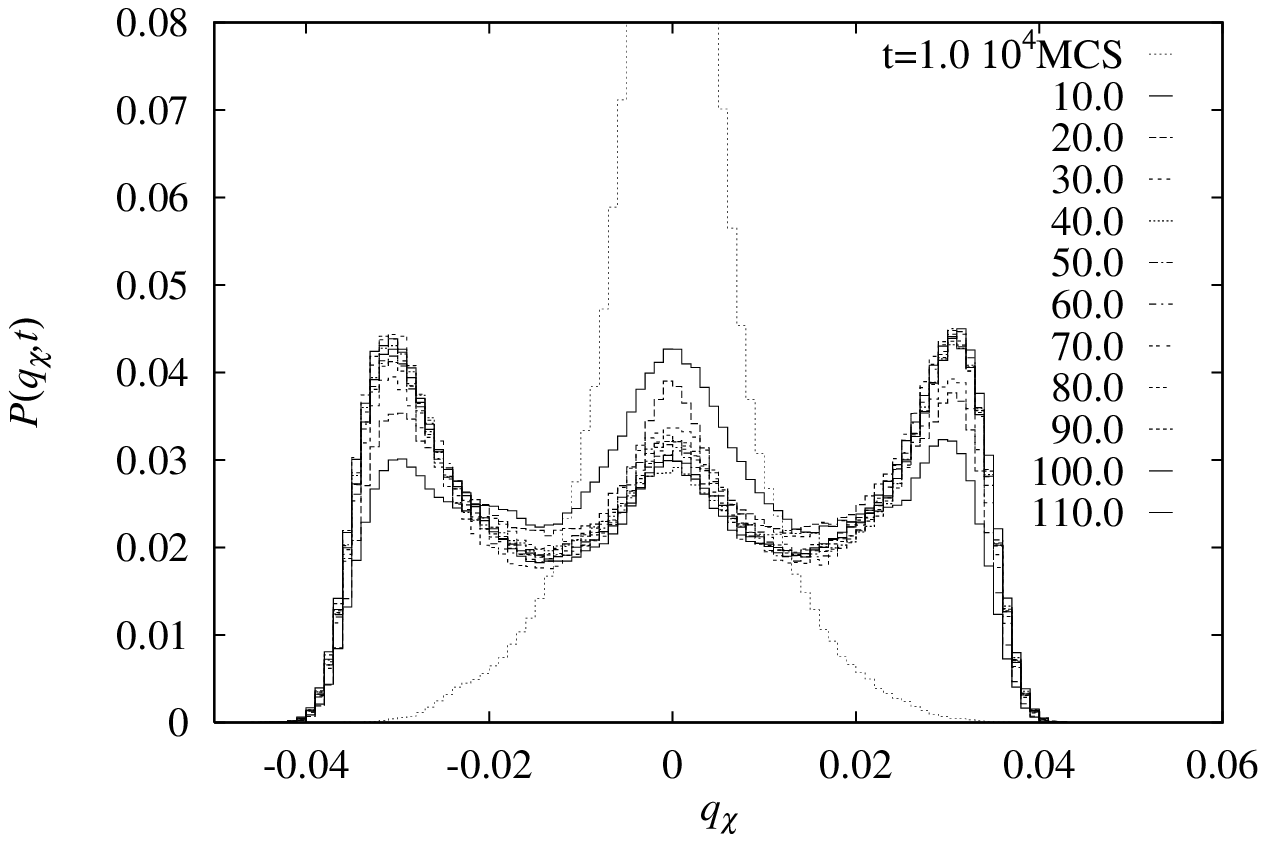}
\epsfxsize=0.48\columnwidth\epsfbox{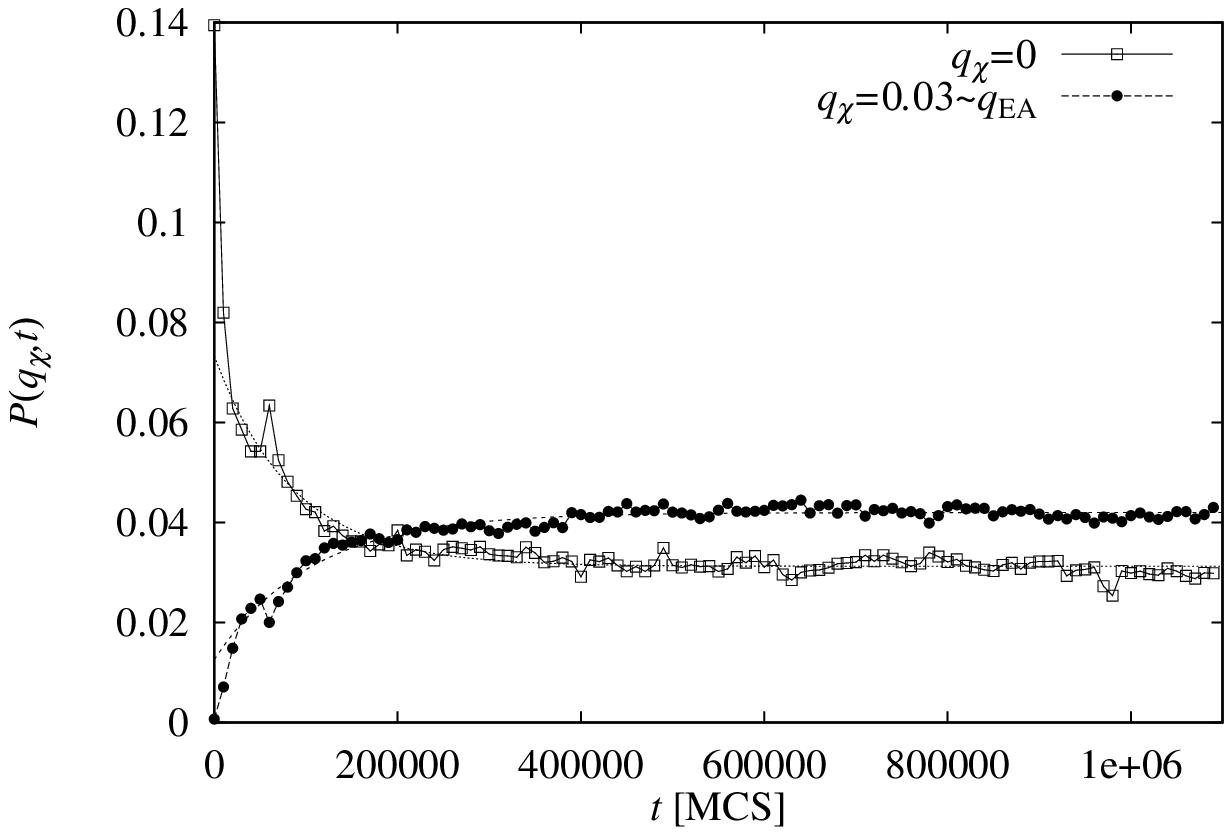}
}
\vspace*{10pt}
\fcaption{
MC-time evolution of the chiral-overlap distribution function at
 $T/J=0.1$ with $L=16$ (left). MC-time dependence of the height of
 $P(q_\chi)$  at $q_\chi=0$ and at  $q_\chi\sim q_{\rm EA}$ (right). 
}
\label{fig:pofq-t}
\end{figure}

Establishing the existence of a finite-temperature chiral-glass transition,
we proceed to the study of the properties of the chiral-glass ordered
state itself.
In Fig.~\ref{fig:pofq}, we display the distribution function of the
chiral-overlap 
defined by
\begin{equation}
P(q'_\chi )=[\langle\delta (q_\chi -q'_\chi )\rangle],
\end{equation}
calculated by the exchange MC method at a temperature $T/J=0.1$,  
well below the 
chiral-glass transition temperature. 
As is evident from Fig.~\ref{fig:pofq}, 
the shape of the calculated $P(q_\chi )$ is
somewhat different from the  one 
observed in the standard Ising-like models such as the
3D EA model or the mean-field
SK model. As usual, 
$P(q_\chi )$ has standard ``side-peaks''
corresponding to the EA order parameter 
$\pm q_{{\rm CG}}^{{\rm EA}}$, which 
grow and sharpen with increasing $L$.
In addition to the side peaks, 
an unexpected  
``central peak'' at $q_\chi =0$ shows up for larger $L$, 
which also grows and sharpens with increasing $L$. This latter
aspect, {\it i.e.\/}, the existence of a central peak growing
and sharpening with the system size, is a peculiar feature of
the chiral-glass ordered phase never
observed in 
the EA or the  SK models. 
Since we do not find any sign of a first-order transition such as a
discontinuity in the energy, the specific-heat nor the order parameter
$q_{{\rm CG}}^{{\rm EA}}$, 
this feature is likely to be related to a 
nontrivial structure in the phase space associated with the
chirality. 
We note that this peculiar feature
is  reminiscent of the behavior characteristic of  some mean-field
models showing the so-called {\it 
one-step replica-symmetry breaking\/} (RSB).\cite{Reviews}
In Figs.~\ref{fig:pofq-t}, we show the (exchange) MC time dependence
of $P(q_\chi )$, together with those of the height of the 
central peak $P(0)$
and of the side peak $P(q_{{\rm EA}})$.
These quantities are
found to reach stationary
values exponentially fast with the correlation time of order of $10^5$
MC steps, 
while actual spin configurations continue to change 
via the temperature-exchange process and the relatively rapid
relaxation realized at higher temperatures.

\setcounter{section}{4}
\setcounter{equation}{0}
\section{Off-equilibrium simulations}
\noindent
In this section, we report on our results of {\it off-equilibrium\/}
MC simulations.\cite{Kawamura98}
Unlike the case of
equilibrium simulations, 
the system here is never in full thermal equilibrium. Recent
studies have revealed that one can still get
useful information from
such off-equilibrium simulations, even including certain
equilibrium properties. The quantities we are mainly 
interested here are
the off-equilibrium spin and  chirality	
autocorrelation functions 
defined by
\begin{eqnarray}
C_s(t_w,t+t_w) & = & \frac{1}{N}
\sum _i[\langle {\mit\bf S}_i(t_w)\cdot {\mit\bf S}_i(t+t_w)\rangle],\\
C_\chi (t_w,t+t_w) & =&\frac{1}{3N}\sum _{i,\mu }[\langle\chi _{i\mu }(t_w) 
\chi _{i\mu }(t+t_w)\rangle].
\end{eqnarray}

MC simulation is performed based on the standard single
spin-flip heat-bath method here.
Starting from completely random initial
configurations, the system is  
quenched to a working temperature.
Total of about $3\times 10^5$ 
MC steps per spin are generated
in each run.
Sample average is taken over 30-120 independent bond realizations, 
four independent runs being made using different
spin initial conditions and different sequences of random numbers
for each sample.
The lattice size  mainly studied is $L=16$ with periodic
boundary conditions, while in some cases lattices with  $L=12$
and $24$ are also studied.
\par\smallskip

The spin and chirality autocorrelation functions 
at a low temperature $T/J=0.05$ are
shown in Fig.~\ref{fig:coft-noneq} as a function of $t$.
For larger $t_w$,
the curves of the spin autocorrelation function  $C_s$ 
come on top of each other
in the long-time
regime, indicating that the stationary
relaxation is recovered and
aging  is
interrupted. This behavior has been expected because the 3D
Heisenberg spin glass has
no standard spin-glass
order.\cite{BanavarCieplak,McMillan,OYS,MII,Kawamura92,Kawamura95,Kawamura96}
By contrast, the chiral 
autocorrelation function $C_{\chi }$ 
shows an entirely different behavior:
Following the initial decay, it exhibits a  clear plateau
around $t\sim t_w$ and then drops sharply for $t>t_w$. 
It also shows an eminent aging effect, namely, as one waits longer,
the relaxation becomes slower and the plateau-like behavior at
$t\sim t_w$ becomes more pronounced. It should be noticed that 
the plateau-like behavior observed here has been 
hardly noticeable
in  simulations of the 3D EA model.

\begin{figure}[h]
\epsfxsize=0.49\columnwidth\epsfbox{fig-noneq-spin.eps}
\epsfxsize=0.49\columnwidth\epsfbox{fig-noneq-chiral.eps}
\vspace*{10pt}
\fcaption{
Off-equilibrium spin (a) and chirality (b) autocorrelation functions
at a temperature $T/J=0.05$
plotted versus 
$\log _{10}t$ for various waiting times $t_w$. The lattice size
is $L=16$ averaged over 66 samples.
}
\label{fig:coft-noneq}
\end{figure}

\begin{figure}[h]
\centerline{
\epsfxsize=0.5\columnwidth\epsfbox{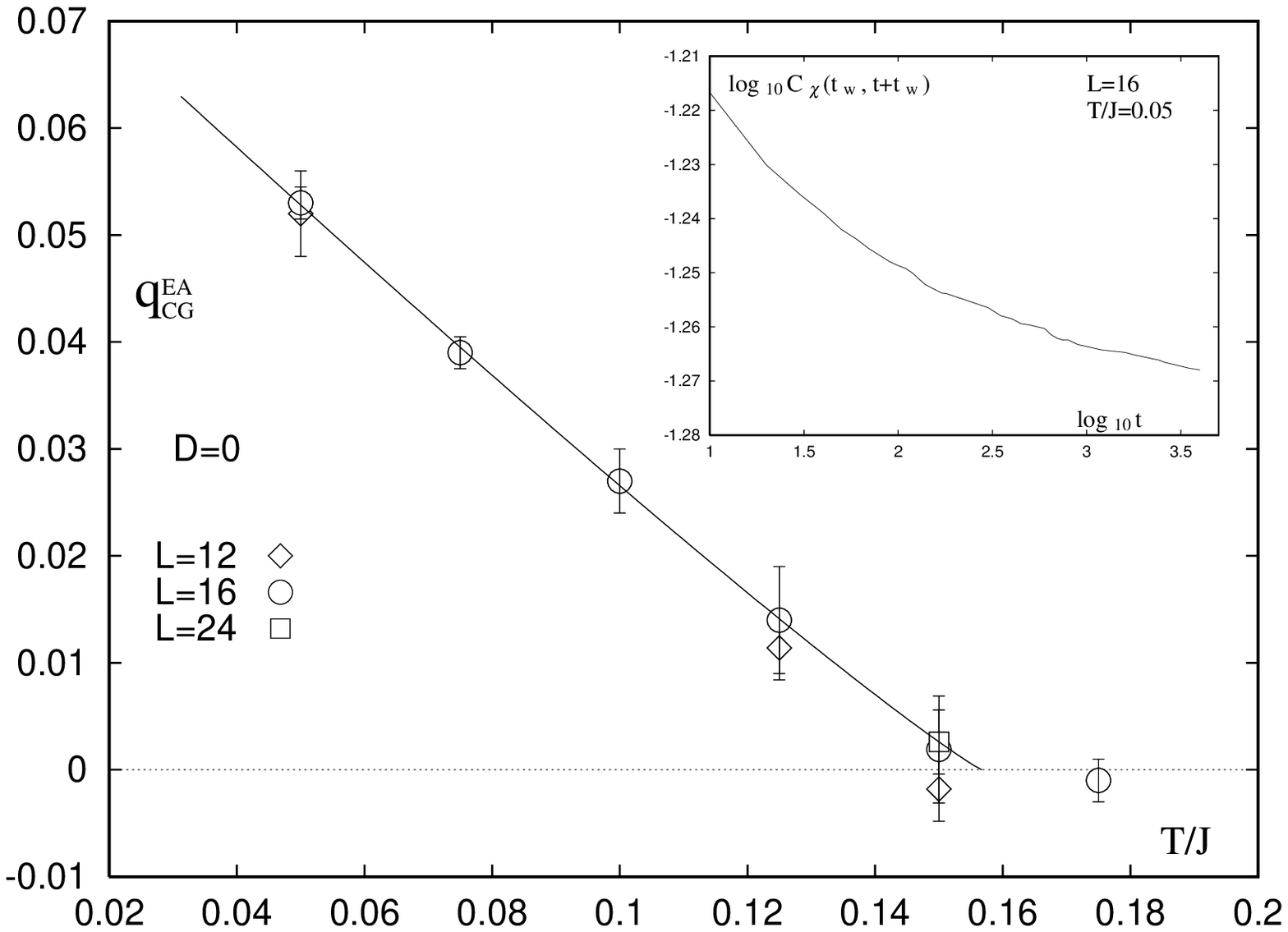}
}
\vspace*{10pt}
\fcaption{
Temperature dependence of the Edwards-Anderson order
parameter of the chirality.
The data are averaged over 30-120 samples.
Inset exhibits the log-log plot of the
$t$-dependence of the
chiral autocorrelation function in the quasi-equilibrium regime
for $L=16$ and 
$t_w=3\times 10^5$. 
}
\label{fig:non-qea}
\end{figure}

While the 
plateau-like behavior observed in $C_\chi $ is already suggestive 
of a nonzero {\it chiral\/} Edwards-Anderson order parameter, 
$q_{{\rm CG}}^{{\rm EA}}>0$,  more quantitative
analysis similar to the one recently done by Parisi {\it et al\/}
for the 4D Ising spin glass\cite{PRR} is performed  
to extract $q_{{\rm CG}}^{{\rm EA}}$ from the data of
$C_\chi $ in the quasi-equilibrium regime.
Finiteness of $q_{{\rm CG}}^{{\rm EA}}$ is also visible in a
log-log plot of
$C_\chi $ versus $t$ as shown in the inset of Fig.~\ref{fig:non-qea},
where the data show a clear upward curvature. 
We extract $q_{{\rm CG}}^{{\rm EA}}$ by fitting 
the data of $C_\chi $ for
$t_w=3\times 10^5$
to the power-law form, 
$C(t_w,t+t_w)=q^{\rm EA}+\frac{C}{t^\lambda}$, 
in the time range
$40\leq t\leq 3,000$
satisfying $t/t_w\leq 0.01$. 
The obtained $q_{{\rm CG}}^{{\rm EA}}$,
plotted as a function of temperature in Fig.~\ref{fig:non-qea},  
clearly indicates the
occurrence of a finite-temperature chiral-glass transition
at $T_{{\rm CG}}/J=0.157\pm 0.01$ with the associated 
order-parameter exponent
$\beta _{{\rm CG}}=1.1\pm 0.1$. 
The size dependence turns out to be rather small,
although the mean values of $q_{{\rm CG}}^{{\rm EA}}$
tend to slightly
increase around $T_{{\rm CG}}$ with increasing $L$.
Since both  finite-size effect and finite-$t_w$ effect
tend to underestimate  $q_{{\rm CG}}^{{\rm EA}}$, one may
regard the present result as a rather strong
evidence of the occurrence of a finite-temperature 
chiral-glass transition. 

The present estimate of the transition temperature 
$T_{{\rm CG}}/J=0.157\pm 0.01$
is in very good
agreement with the equilibrium estimate in the preceding section, 
$T_{{\rm CG}}/J=0.160\pm 0.005$.
If we combine the present estimate of $\beta _{{\rm CG}}$ with
the estimate of $\gamma _{{\rm CG}}$ in the preceding section and
use the scaling relations, 
various chiral-glass exponents can be estimated to be
$\alpha \simeq -1.7$, $\beta_{{\rm CG}}\simeq 1.1$,
$\gamma_{{\rm CG}}\simeq 1.5$,  $\nu_{{\rm CG}}\simeq 1.2$ and
$\eta _{{\rm CG}}\simeq 0.8$. 
The dynamical exponent is estimated to be $z_{{\rm CG}}\simeq 4.7$ 
by using the estimated value of $\lambda $
and the scaling relation $\lambda =\beta _{{\rm CG}}/z_{{\rm CG}}
\nu _{{\rm CG}}$. While
the dynamical exponent $z_{{\rm CG}}$ comes rather 
close to the $z$ of the 3D Ising EA model,
The obtained static exponents differ significantly
from those of the 3D Ising EA model
$\beta\simeq 0.6$,
$\gamma\simeq 4$,  $\nu\simeq 2$ and $\eta \simeq
-0.35$,\cite{KawashimaYoung,HTN,MPR,BergJanke,Caracciolo} 
suggesting that the universality class of the chiral-glass
transition of the 3D Heisenberg spin glass 
differs from that of the standard 3D Ising spin glass. 
According to the chirality mechanism,
the criticality of real
spin-glass transitions 
is derived from that of the chiral-glass
transition of an isotropic Heisenberg spin glass, so long as the
magnitude of  random anisotropy is not too strong.
If one tentatively accepts this scenario, 
the present result opens up an interesting possibility
that the universality class of many of real spin-glass transitions 
might differ from that of the standard Ising
spin glass, contrary to common belief.

\section{Conclusion}
\noindent
In summary, spin-glass and chiral-glass orderings of 
isotropic 3D Heisenberg spin glasses are studied 
by extensive MC simulations. 	
Clear evidence of the occurrence of a 
finite-temperature chiral-glass transition without the
conventional spin-glass order
is  presented both by equilibrium and off-equilibrium simulations.
Spin and chirality
show very different dynamical behaviors consistent with the
``spin-chirality separation''. While the spin 
autocorrelation exhibits only
an interrupted aging, the chirality autocorrelation
persists to exhibit a pronounced aging
effect reminiscent of the one observed
in the mean-field model. The universality
class of the chiral-glass transition is 
different from that of the the standard
Ising spin glass, while the chiral-glass ordered state appears
to exhibit a feature of ``one-step'' replica-symmetry breaking.
We
expect that these
numerical findings have important implications to the
understanding of the nature of 
real spin-glass ordering.

\nonumsection{Acknowledgment}
\noindent
We would like to thank Prof. I.~Campbell, Prof. A.~P.~Young and
Prof. H.~Takayama for useful discussion. 
The numerical calculation was performed on the Fujitsu VPP500 at the
supercomputer center, ISSP, University of Tokyo, on the Hitachi SR2201
at the supercomputer center, University of Tokyo, and on the CP-PACS
computer at the Center for Computational Physics, University of
Tsukuba. 

\nonumsection{References}

\end{document}